\documentstyle[12pt]{article}

\begin{document}

\title{The Thermodynamics of Quantum Systems and Generalizations of
Zamolodchikov's C-theorem}

\vspace*{1.0cm}

\author{A.~H.~ Castro Neto and Eduardo Fradkin \\
Department of Physics \\
University of Illinois at Urbana-Champaign \\
1110 W. Green St., Urbana, Illinois 61801-3080}

\vspace{0.5cm}

\date{October 23, 1992}

\maketitle

\vspace{0.8cm}

\setcounter{page}{1}

\begin{abstract}
In this paper we examine the behavior in temperature of the free
energy on quantum systems in an arbitrary number of dimensions.
We define from the free energy a function $C$ of the coupling
constants and the temperature, which in the regimes where quantum
fluctuations dominate, is a monotonically increasing function
of the temperature. We show that at very low temperatures the system
is controlled by the
zero-temperature infrared stable fixed point while at intermediate
temperatures the behavior is that of the unstable fixed point. The
$C$ function displays this crossover explicitly.
This behavior is reminiscent of Zamolodchikov's
$C$-theorem of field theories in 1+1 dimensions. Our results are
obtained through a thermodynamic renormalization group approach. We
find restrictions on the behavior of the entropy of the system for a
$C$-theorem-type behavior to hold. We illustrate our ideas in the
context of a free massive scalar field theory, the
one-dimensional quantum Ising Model and the quantum Non-linear Sigma
Model in two space dimensions. In regimes in which the classical
fluctuations are important the monotonic behavior is absent.
\end{abstract}

\vspace{0.7cm}

\newpage

\section{Introduction}

\indent

Our understanding of the behavior of systems near critical points in
1+1 dimensions has been
significantly improved in recent years due to the application of
the methods of conformal field theory
\cite{bela}. One of the aspects of this method which makes it
particularly powerful and elegant is that it relates in a simple way
the microscopic behavior of such systems with thermodynamics.
At the critical point the most important
parameter of the theory is the central charge, $c$, which is related
with the quantum fluctuations at zero temperature. The central charge
can be calculated in many ways. At zero temperature it can be calculated
from the correlation function of the energy momentum tensor or from
the commutation relations between the generators of the Virasoro
algebra \cite{cardy}. It can be shown that $c$ can also be obtained
at finite temperatures from the specific heat of the system and,
therefore, it can be measured in an experiment.

Conformal Field Theory (CFT) describes the behavior of systems {\it at }
fixed points. Away from fixed points, CFT can also describe the ``early"
an ``late" stages of the crossover between fixed points. The
behavior at intermediate scales is, in general, non-universal and
thus depends on the microscopic description of the system. The
universal behavior of this crossovers is described by
the famous Zamolodchikov's $C$-theorem \cite{Zamol}. This theorem
states that it is possible to define a function $C$ of the coupling
constants of the theory which decreases monotonically  under
the renormalization group flow, that is, as we change the characteristic
scale
the function $C$ goes monotonically from a fixed point of larger
central charge to one at smaller central charge. The assumptions present
in his proof are the euclidian invariance, the existence of a symmetric,
conserved, energy momentum tensor and the unitarity of the field theory.
The physical interpretation of this theorem is related with loss of
information due to the coarsed graining procedure
present in the renormalization group equations. Another way to
interpret this result asserts that $C$ measures the number of
fluctuating degrees of freedom. These must decrease when we
integrate over small scales in the renormalization group flow.

All the results of the conformal field theory seem to be very
difficult to be applied for higher dimensions due to the proliferation
of components in the energy momentum tensor. Nevertheless some
importants attemps have been made with the use of the spectral
representation for the correlation function of the energy momentum
tensor \cite{cappelli1}, \cite{cappelli2}. However, the relation between
these results and thermodynamics remains unclear up to now.

The approach of this paper is based on the behavior of the
thermodynamic functions under the renormalization group flow.
Following this route we can determine the macroscopic
conditions
which are necessary for any system, in any number of dimensions,
to have a behavior which is analogue to the one described by
the $C$-theorem proposed by Zamolodchikov. We can show that the
the monotonic behavior implied by the $C$-theorem can not be derived
from the laws of thermodynamics alone
and, therefore, this behavior is not universally true.
As we will show, the sort of behavior needed for a $C$-theorem to hold
is only possible in regimes in which quantum fluctuations dominate. In
1+1 dimensions this is always the case because of the triviallity of
one-dimensional classical statistical mechanics. But, in higher
dimensions, there is a complex quantum-to-classical crossover which
spoils the general validity of a $C$-theorem. To put it differently,
this is a consequence of the fact that 1+1 dimensions {\it all} systems
are {\it at} or {\it below} their lower critical dimension. This fact
imposes stringent constraints on the allowed behavior of the entropy of
these systems at low temperatures. However, in higher dimensions a
variety of complex behaviors are allowed and the constraints on the
entropy are much weaker. We believe that this observation is a hint that
there is no general $C$-theorem in dimensions higher than 1+1.

We have investigated these issues in the context of several systems.
We first consider the case of a free massive real scalar field in
arbitrary dimensions. The results also apply to more general
bosonic systems as well as to the case of relativistic fermions. Here,
the crossover parameter is the mass. We find monotonic behavior
in any number of dimensions.
We also discuss two more non-trivial
models. The first is the quantum Ising model in 1+1 dimensions.
We show that the the monotonicity of the $C$-function (calculated
without approximations) defined from
the themodynamics is obeyed in the low temperature limit and that
near the scaling limit (where the model can be described as a
field theory of free massive fermions) we found all the results expected
from the point of view of conformal field theory. At temperatures
comparable or higher than the fermion ``bandwidth", there is a crossover
to classical behavior and monotinicity is lost (this is the ``Debye
temperature"). Next we discuss
the example of the Non-linear Sigma model (the continuum version of the
quantum  Heisenberg antiferromagnet) in 2+1 dimensions.
We show that the $C$-function (calculated from the $1/N$ expansion) is
not monotonic for arbitrary temperatures and coupling constant due to a
crossover
from classical to quantum behavior. This crossover can be observed
even near the phase transition. Once again, in the quantum regime we
find  a term in the free energy which displays the same monotonic
behavior expected from a $C$-theorem.

In order to understand these results, we observe that
temperature plays two different roles
in these problems. At low temperatures, these systems are essentially
quantum in $d+1$ dimensions. Here the temperature mainly affects the
geometry in which the system lives since it measures the ``perimeter" of
Euclidean space-time in the (imaginary) time direction. As the
temperature is raised, this perimeter grows smaller and the temperature
drives the system away from criticallity, just like a relevant operator.
On the other hand, at high temperatures the
dimension in the imaginary direction effectively shrinks to zero
(compared with all the length scales in the system) and the
system behaves like a classical system in $d$ dimensions. In this case
the
temperature only plays the role of a coupling constant in classical
statistical mechanics. Thus, at these high temperatures, the $C$-function
of the system is not necessarily monotonic. For instance, one does not
expect monotonicity
if the classical system has a phase transition at a non zero
temperature. Even if this is not the case, such as in classical systems
at or below their lower critical dimension, non-monotonicity may still
occur if the classical system has gapless configurations. We will see
below that this is the case of the Non-linear Sigma Model in the
``classical regime". In the case of systems with discrete symetries
({\it e.g.,} the Ising Model), non-monotonicity still occurs near what we
will refer to as the classical fixed point, but not near the critical
zero-temperature fixed point. Thus, we do not expect to have a monotonic
behavior of any piece of the free energy {\it for all} temperatures but
only in regimes in which the quantum fluctuations are dominant. The
strategy of this paper is thus to look for the regimes of parameter
space ( {\it i.e.} coupling constants and temperature) in which the
physics is dominated by the quantum fluctuations.

In the next section we discuss the relation between the Renormalization
Group of quantum systems at finite temperature and Thermodynamics.
In section 3 we discuss our results in the context of free scalar and
fermi fields.
In section 4 we apply our
ideas for the case of the quantum Ising model in 1+1 dimensions. In
section 5 we show how
our results work for the case of the Non-linear Sigma model in 2+1
dimensions. The last section contains our conclusions.

\section{The Renormalization Group and Thermodynamics}

\indent
Suppose we are interested in the study of a system in statistical
mechanics which can be described in a lattice with some characteristic
length $a$ (UV cut-off) and which has a set of dimensionful coupling
constants ${g}$. By naive dimensional analysis we know that the bulk free
energy of this system in thermal equilibrium at temperature $T$ ($\beta=
1/T$, $\hbar=k_B=1$) in $(d+1)$-dimensions, in the thermodynamic limit,
can be written as
\begin{equation}
F(\beta,g,a)=E_0(g,a)-\frac{n(d) V C(\beta,g,a)}{v^d \beta^{d+1}}
\end{equation}
where $E_0$ is the zero temperature energy, $V$ is the volume
($V \to \infty$, but $N/V$ is finite, where N is the number of particles),
$v$ is a characteristic velocity in the system (for instance, the velocity
of the quasi particles), $n(d)$ is a positive real number which depends
only on the dimensionality of the system (see next section for details)
and $C$ is a dimensionless function of its arguments.

We want to comment here that in (2.1) we are implicitly assuming that
the ultra-violet divergences in the field theory can be subtracted in
the first term on the r.h.s. of (2.1) since these divergences only occur
at zero temperature, that is, since $C$ is a finite size effect (in terms
of temperature) it is insensitive to divergences which occur in small
scales \cite{boya}.

Here we are interested in the limit when $a \to 0$, that is, in the limit
where the system can be described by a field theory (we do not intend to
discuss here how this can be achieved, we just assume that is possible).
In this limit the ``bare" coupling constants of the original theory are
replaced by renormalized dimensionless coupling constants ${\alpha}$
and since $C$ is dimensionless we should write
\begin{equation}
C(\beta,g,a)=C(\alpha,\beta \Lambda)
\end{equation}
where $\Lambda$ has dimensions of inverse of length.

Since $C$ is an observable it does not change as the scale is changed and
therefore it obeys a renormalization group (RG) equation ($t=(\beta
\Lambda)^{-1}$)
\begin{equation}
\left(t \frac{\partial}{\partial t} -\tilde{\beta}(\alpha) \frac{\partial}
{\partial \alpha}\right) C(\alpha,\beta\Lambda)=0
\end{equation}
where
\begin{equation}
\tilde{\beta}(\alpha)=\Lambda \left(\frac{\partial \alpha}{\partial \Lambda}
\right)_{g,\beta}
\end{equation}
is the beta function of the system. Observe that the beta function
can also be obtained from (2.3).

The solution of (2.3) is obtained with the introduction of the correlation
length of the system, $\xi(\alpha,t)$, defined by its renormalization group
equation \cite{zinn}
\begin{equation}
t \, \frac{\partial \xi(t)}{\partial t}=\tilde{\beta}(\alpha)
\frac{\partial \xi}{\partial \alpha}
\end{equation}

It is trivial to show that
\begin{equation}
C(\alpha,t)=C\left(\frac{\xi(\alpha,t)}{\beta}\right).
\end{equation}

Define now a dimensionless correlation length by
\begin{equation}
\gamma = \Lambda \xi
\end{equation}
and therefore from (2.6)
\begin{equation}
C \, = \, C( t \, \gamma(\alpha,t))
\end{equation}
and from (2.5) we have
\begin{equation}
\frac{\partial(t \gamma)}{\partial t}=\tilde{\beta}(\alpha)
\frac{\partial \gamma}{\partial \alpha}.
\end{equation}

It is also important to define the running coupling constant, $\alpha(t)$,
as the solution of the equation (2.4)
\begin{equation}
t \frac{\partial \alpha(t)}{\partial t}=\, - \, \tilde{\beta}\left(
\alpha(t) \right)
\end{equation}
with the initial condition
\begin{equation}
\alpha\left(t=t_0\right)=\alpha
\end{equation}
where $t_0$ is an arbitrary renormalization point.

It is well known that at a critical point (where the conformal field theory
can be applied \cite{cardy}) the function $C$ as defined in (2.1) is a
constant and, in one dimension, is related with the Virasoro algebra
(it is called the central charge of the theory).

Zamolodchikov \cite{Zamol} proposed some years ago a function $C_Z(R,g)$
(where $R$ is a length scale in the 1+1 dimensional space) which is
monotonically decreasing under the RG flow, that is,
\begin{equation}
R \left(\frac{\partial C_Z}{\partial R}\right)_{\alpha} \leq 0
\end{equation}
and which, at the critical point, is the central charge. His proof is
based on the euclidian invariance, the existence of a symmetric, conserved,
energy-momentum tensor and unitarity of the field theory.

Suppose we can actually calculate the spectrum for the system under
consideration. Of course the free energy can always be written in the
form (2.1) but, in this case, since $C$ is dimensionless,
\begin{equation}
C=C(\phi(g) \beta)
\end{equation}
where $\phi(g)$ is a characteristic energy scale which depends on
the bare coupling constants of the theory and possibly on the temperature.
In general we are talking about a Hamiltonian with the form
\begin{equation}
H=H_0 + g H_1
\end{equation}
where $H_0$ is the fixed point Hamiltonian and $H_1$ is a ``pertubation"
which drives the system away from criticallity. In this case when $g$
vanishes $C$ goes to the central charge at the fixed point,
\begin{equation}
Lim_{g \to 0} C(\phi(g) \beta)= c_0 = constant.
\end{equation}

Since in thermodynamics, for each
system, the coupling constants are fixed, the only energy scale free
to be changed is the temperature. Therefore we proposed an analogue of the
Zamolodchikov`s equation (2.12) which is valid in any number of dimensions
\begin{equation}
\beta \left(\frac{\partial C}{\partial \beta}\right)_{\alpha} \leq 0
\end{equation}

Observe that we are using the fact that in finite temperature field
theory we have to impose periodic (or anti-periodic) boundary conditions
in the imaginary time direction. These boundary conditions permit
us to interpret a quantum system in $d+1$ dimensions as a classical
system in $D (=d+1)$ dimensions. The manifold under consideration
has the form of a cilinder since the other $d$ dimensions are free.
The length of the cilinder in the imaginary time direction is $\beta$
and, therefore, (2.12) and (2.16) represent the same change in the scale,
although, we could say, (2.12) is a isotropic change while (2.16) is
anisotropic. Observe that at high temperatures the length of the
cilinder goes to zero and therefore the field theory is restricted
to $d$ dimensions.

{}From the point of view of the renormalization group approach we see
that if we follow the flow using the running coupling constant
defined by equation (2.10) the function C is stationary as
required. However if we fix the renormalized coupling constant
(as required in Zamolodchikov's theorem) and vary the
dimensionless temperature, $t$, we would get (2.16).
Using the result (2.8) and the RG equation (2.9) we find,
\begin{equation}
\frac{\partial C}{\partial t}=\frac{d C}{d(t\gamma)}
\, \frac{\partial(t\gamma)}{\partial t}=\frac{d C}{d (t\gamma)}
\, \frac{\partial \gamma}{\partial \alpha} \, \tilde{\beta}(\alpha)
\end{equation}
This expression will be useful later.

Consider a system initially at temperature zero in a volume
$V$, with ground state energy $E_0$ and entropy $S_0$. Suppose we
perform a thermodynamic transformation at a constant volume which
increases the temperature of the system up to a temperature $T$.
At this point the system, in thermal equilibrium, has an internal
energy $U$ and an entropy $S$. We will show that if,
\begin{equation}
\left(\frac{d+1}{d}\right) \frac{\delta U}{T} \geq \delta S
\end{equation}
where
\begin{equation}
\delta U=U-E_0
\end{equation}
and
\begin{equation}
\delta S=S-S_0
\end{equation}
then (2.16) follows. We will show at the end of this section
that (2.18) can not be obtained from the laws of thermodynamics.
Therefore the irreversibility of the RG flow expressed
in (2.16) is a consequence of (2.18).

Notice that (2.18) has a very nice interpretation
in terms of the number of states. Since in the thermodynamical limit
the entropy of a system can be writen as \cite{reif}
\begin{equation}
S(U)=\ln(\Omega(U))
\end{equation}
where $\Omega(E)$ is the thermodynamic number of states defined as
the number of states with energy between $E$ and $E+dE$, we can
reexpress (2.18) as
\begin{equation}
\Omega(E_0) e^{\frac{d+1}{d} \frac{\delta U}{T}} \geq \Omega(U)
\end{equation}
which means that the number of states with energy between $U(T)$ and $U(T)
+dU(T)$ is bounded from above. This is a highly
non trivial statement which, as we shall see below, is not required by
the laws of thermodynamics alone.

Before we start to prove that (2.16) is a consequence of (2.18)
let us remember that the continuum limit $a \to 0$
is very important in Zamolodchikov`s proof. There, he assumed that,
\begin{equation}
R \gg a
\end{equation}

In this paper we also assume the counterpart of (2.23) in terms of
temperature, namely,
\begin{equation}
T \ll \theta
\end{equation}
where $\theta$ is a characteristic cut-off temperature
(for instance, the Debye temperature, $\theta_D$).

Using equation (2.1) the thermodynamical quantities, pressure ($P$),
entropy ($S$) and thermal energy ($U$) can be easily calculated,
\begin{equation}
P=\frac{n(d) C}{v^d \beta^{d+1}}
\end{equation}
\begin{equation}
S=\frac{n(d) V}{v^d \beta^d}  \left((d+1)C - \beta \frac{\partial C}{\partial
\beta}\right)
\end{equation}
\begin{equation}
U=E_0 + \frac{n(d) V}{v^d \beta^{d+1}} \left(d \, C - \beta \frac{\partial C}
{\partial \beta} \right)
\end{equation}

Notice that as the temperature goes to zero the free energy and the
internal energy goes to the ground state energy. This fact means that
$C$ and $\partial C/ \partial \beta$ go to zero in this limit. From
the third law of thermodymanics we know that
\begin{equation}
Lim_{T \to 0} S(T)=S_0
\end{equation}
So in order to our definition of $C$ be consistent we have to set
\begin{equation}
S_0=0
\end{equation}
which is allowed from the thermodynamic point of view.

{}From (2.26) and (2.27) we can write
\begin{equation}
\beta \frac{\partial C}{\partial \beta} = \frac{v^d \beta^{d}}{n(d) V}
( d \, \, S - (d+1) \beta \delta U)
\end{equation}
therefore if (2.18) is obeyed (2.16) follows and we get the same monotonicity
propertie as in the original C-theorem.

We can also rewrite (2.30) in an useful form. Recall that the definition of
the specific heat at constant volume is
\begin{equation}
C_v(T)= T \frac{\partial S}{\partial T}= \frac{\partial U}{\partial T}
\end{equation}
which, when integrated using (2.29), gives
\begin{equation}
S(T)=\int_{0}^{T} \frac{C_v(T')}{T'} dT'
\end{equation}
and
\begin{equation}
\delta U(T) = \int_{0}^{T} C_v(T') dT'.
\end{equation}

Substituting (2.32) and (2.33) in (2.30) we find
\begin{equation}
\beta \frac{\partial C}{\partial \beta} = - \frac{v^d \beta^d}{n(d) V}
\int_{0}^{T} C_v(T') \left(\frac{1+d}{T}-\frac{d}{T'}\right) dT'
\end{equation}
Expression (2.34) is very suitable for experimental test and can be used to
classify the systems which obey assumption (2.18). Morover, at the fixed point
we expect that $\partial C/ \partial \beta = 0$ and therefore the specific
heat must behave like $T^d$ which is known to be true in one dimension.

Now we can explain why condition (2.24) is important for our results.
The pertubation term in (2.14) breaks the scale invariance of
the fixed point hamiltonian and it drives the system into the basin of
attraction of a more stable fixed point. Once temperature is turned on,
this operator generates flows away from the fixed point but which are
``cutoff" once
the correlation length $\xi$ is of the order of the inverse temperature.
Thus, at temperatures higher than $\xi^{-1}$, the system behaves as if
it were still at the infrared unstable fixed point but, at lower
temperatures it crosses over to the infrared stable one. It is this
crossover which is described by Zamolodchikov's theorem \cite{ludwig}.
Morover, at high
temperatures the specific heat must go to its classical value which is
a constant given by the equipartition theorem (Dulong-Petit law),
therefore, the entropy diverges logarithmically and we do not expect that
assumption (2.18) is obeyed. However, if the spectrum has a width
which is much
greater than the energy scale $\phi(g)$ we could imagine a situation
where the temperature is much greater than $\phi(g)$ but much smaller
than the bandwidth. In this case, since $C$ is only a function of
$\phi(g) \beta$, the limit as the temperature goes to infinity
is identical (in terms of RG) to the limit as $g \to 0$ and therefore,
by (2.15), we see that the $C$ is the central charge in this point.
This is a stable high temperature fixed point.
Nevertheless,
for temperatures larger than the bandwidth we expect that the behaviour
of the system is classical and (2.18) will not be obeyed.
We will see later that this behaviour can be seen in the Ising model
in 1+1 dimensions.

For systems with a phase transition (when $\phi$ also
depends on the temperature) we know that near the critical temperature
the correlation length of the system increases exponentially and therefore
we do not expect that the thermodynamic number of states be bounded from above.
This
behaviour near the critical temperature is essentially classical and therefore
the monotonicity of $C$ should be violated. We will show that this is what
happen in the Non-linear Sigma Model.

We want to point out that in the assumption that we can describe the
system by a quantum field theory we are assumig that there is a zero
temperature
fixed point which is unstable under the increase of the temperature. This is
clearly expected in any quantum system since we always have the crossover
from the quantum to the classical system as the temperature increases.
Therefore, (2.16) shows that the RG's flow goes from the zero temperature
fixed point to a fixed point at finite temperature ( in particular to a
high temperature fixed point).

In order to finish this section  we will show that we can not obtain
(2.18) from the laws of thermodynamics.

Suppose we have initially the system at zero temperature and we put it in
contact with a thermal reservoir at temperature $T$ and wait for the
thermal equilibrium. Suppose that this system is isolated from the
rest of the universe. The variation of the entropy of the reservoir is
$-Q/T$ where $Q$ is the heat given by the reservoir to the system.
Using the second law we get
\begin{equation}
S \geq \frac{Q}{T}
\end{equation}

Using the first law we have
\begin{equation}
\delta U = Q
\end{equation}
therefore
\begin{equation}
S \geq \frac{\delta U}{T}
\end{equation}
We see that the laws of thermodynamics can not lead to (2.18), that
is, they show that the increase of the entropy is bounded from
below.
Furthemore, notice that from (2.37), (2.26) and (2.27) we obtain that
$C$ is always positive as required.

\setcounter{equation}{0}

\section{Free Massive Field Theories}

We will now discuss the physical meaning of the results of the last
section in the context of specific systems.
We will first consider the case of a massive scalar field at
non-zero temperature and exhibit the crossover. We will do that by
showing that the restriction (2.18) is always obeyed.

The approach is very straightfoward. The free energy for a system
with one-particle excitation spectrum $E_{\kappa}$ where $\kappa$
is a set of quantum numbers is given by \cite{reif}
\begin{equation}
F=E_0 + \frac{\tau}{\beta} \sum_{\kappa} \ln \left(1-\tau e^{-\beta E_{\kappa}}
\right)
\end{equation}
where $\tau$ is +1 (-1) for bosons (fermions).

Comparing (2.1) and (3.1) we get
\begin{equation}
C=-\frac{(v \beta)^d}{n(d) V} \tau \sum_{\alpha} \ln\left( 1 -
\tau e^{-\beta E_{\alpha}} \right)
\end{equation}

Suppose we have the relativistic dispersion relation
\begin{equation}
E_{k}=\sqrt{m^2 v^4 + v^2 k^2}
\end{equation}
where $k$ is the wavenumber (in units of $\hbar$) which is defined
by periodic boundary conditions. In the thermodynamic limit we can
replace the sum in (3.2) by an integral as usual. We will also assume
$v=1$ and rewrite (3.2) as
\begin{equation}
C=-\frac{\beta^d}{n(d) (2 \pi)^d} \tau \int d^dk \, \, \ln \left( 1-
\tau e^{-\beta E(k)}\right)
\end{equation}
since the spectrum only depends on the absolute value of $\vec{k}$, we have
\begin{equation}
C=-\frac{\beta^d}{n(d) 2^{d-1} \pi^{d/2} \Gamma(d/2)}
\tau \int_{0}^{\infty} dk \, k^{d-1} \, \ln\left(1-\tau e^{-\beta E(k)}\right)
\end{equation}
where $\Gamma(x)$ is the Gamma function.

Using (3.3), changing variables in the integral (3.5), expanding the logarithm
in powers and integrating, we easily get
\begin{equation}
C( m \beta)=\frac{(m \beta)^{\frac{d+1}{2}}}{n(d) 2^{\frac{d-1}{2}}
\pi^{\frac{d+1}{2}}}
\sum_{l=1}^{\infty} \frac{\tau^{l+1}}{l^{\frac{d+1}{2}}} K_{\frac{d+1}{2}}
(l \, m \beta)
\end{equation}
where $K_{\nu}(x)$ is the Bessel function of imaginary argument of order
$\nu$.

Observe that, as expected, $C$ is only function of one variable, namely
$m \beta$, and, as explained before, we find that the correlation length
is given by $m^{-1}$. Furthermore, since $\alpha=\Lambda^{-1} m$ (
$\gamma=\alpha^{-1}$), the
beta function is trivial, namely, $\tilde{\beta}(\alpha)=-\alpha$, and
the running coupling constant as defined in (2.10) is just $\alpha(t)
= \alpha t/t_0$.

We can also obtain the special cases. In the massless limit, when kinetic
energy of the system
(given by $m$) is much smaller than thermal energy (given by $\beta^{-1}$)
that is, $ m \beta << 1$ ($\xi >> 1/T$), we can approximate the Bessel
function for small argument \cite{grads} and write
\begin{equation}
C(0)=\frac{\Gamma\left(\frac{d+1}{2}\right)}{n(d) \pi^{\frac{d+1}{2}}}
\sum_{l=1}^{\infty} \frac{\tau^{l+1}}{l^{d+1}}
\end{equation}
which is finite. This is the high temperature fixed point.

{}From (3.6) and using well know relations between Bessel functions
\cite{grads} we find that the rate of change of $C$ with the temperature
is given by,
\begin{equation}
\frac{d C}{d \beta}= -\frac{2 m}{n(d)} \left(\frac{m \beta}{2 \pi}\right)^
{\frac{d+1}{2}} \sum_{l=1}^{\infty} \frac{\tau^{l+1}}{l^{\frac{d-1}{2}}}
K_{\frac{d-1}{2}}(l m \beta)
\end{equation}
which is always negative in any number of dimensions. We could also
arrive to this result using equation (2.17).

However the normalization factor $n(d)$ is still arbitrary.
For relativistic systems in 1+1 dimensions, this factor, at a fixed
point, equals to the central charge of the Virasoro algebra which can be
thought as a measure of the number of fluctuating fields. Its well
known \cite{affleck} that in 1+1 dimensions relativistic massless bosons
have $C(0)=1$ and relativistic massless spinless fermions
(Majorana fermions) have $C(0)=1/2$. Hence,
\begin{equation} n(1)=\frac{\pi}{6}.
\end{equation}

In higher dimensions we
do not know if a generalization of the notion of central
charge exists. Still, we may {\it choose} the normalization
factor in such a way that, for free fields, it counts the
number of degrees of freedom. This choice leads to the {\it
definition}
\begin{equation}
n(d)=\frac{\Gamma\left(\frac{d+1}{2}\right)}{\pi^{\frac{d+1}{2}}}
\zeta(d+1)
\end{equation}
for bosons and
\begin{equation}
n(d)=\frac{\Gamma\left(\frac{d+1}{2}\right) \left(2^d-1\right)}
{\pi^{\frac{d+1}{2}} 2^{d-1}} \zeta(d+1)
\end{equation}
for fermions, where $\zeta(x)$ is the Riemman's Zeta function. Observe
that spinless fermions and bosons only have the same $n(d)$ at $d=1$.
Anyway this arbitrariness in $n(d)$ does not
affect our discution about the mononicity properties of the
$C$ function.

Let us remark that the behaviour of the function $C$ is non
analitic as a function of $m$, that is, if $m$ is strictly zero
the function $C$ is a constant given by (3.7) in all ranges of
temperatures. If $m$ finite,
even very small, the behaviour of $C$ is completely different
(see Fig.1).
Indeed, in the massive limit, $m \beta >> 1$ ($\xi << 1/T$),
or low temperatures,
the Bessel function in (3.6) vanishes exponentially and $C$ is zero.

Finally, for systems like Fermi liquids we know that independent of the
dimensionality the specific heat is linear in the temperature \cite{pines}.
If we substitute this result in (2.34) we will see that $\partial C/
\partial \beta$ is negative in two or three dimensions but it is
positive in one dimension. This result is clear from the point
of view of relativistic particles since in one dimension we can
always expand the energy around the Fermi surface as in (3.3).

\setcounter{equation}{0}

\section{The Ising Model}

Here we will consider the Ising model which can be described by the
following Hamiltonian
\begin{equation}
H = - \Gamma \sum_{n=1}^{N} S_{n}^{3} - J \sum_{n=1}^{N} S_{n}^{1}
S_{n+1}^{1}
\end{equation}
where $S_{n}^{j}$ is the $j^{th}$ ($j=1,2,3$) projection of the spin operator
in
the site $n$ ($n=1,...,N$). We will assume periodic boundary conditions,
that is,
\begin{equation}
S_{N+1}^{1}=S_{1}^{1}
\end{equation}
and neglect any surface effect which might occur.

This model was solved exactly \cite{lieb} using the Jordan-Wigner
transformation. Since the procedure is somewhat standard we will
only quote the results (see \cite{lieb} for details).

It is possible to show that the problem reduces to a problem
of fermions with dispersion relation given by
\begin{equation}
E_k = \Gamma \sqrt{1 \,+ \, \varrho^2 \, - \, 2 \, \varrho \,
\cos(k)}
\end{equation}
where
\begin{equation}
k=\frac{2 \pi m}{N}
\end{equation}
with $m=0,\pm1,\pm2,...,\pm \frac{N}{2}$ ($N  \to  \infty$).
The constant $\varrho$ is defined as
\begin{equation}
\varrho = \frac{J}{2 \Gamma}
\end{equation}
and it controls the physics of the problem. Observe that when
$\varrho = 1$ the dispersion relation is simply
\begin{equation}
E_k = 2 \, \Gamma \, \sin\left(\frac{k}{2}\right)
\end{equation}
and the fermions are massless. The velocity of the excitations
can be easily computed to be $\Gamma$ near $k=0$. At this point
the system is scale invariant and it represents the phase transition.

The free energy is obtained straightfowardly \cite{pfeuty}
\begin{equation}
F=E_0 - \frac{N}{\beta \pi} \int_{0}^{\pi} dk \, \ln\left(1+e^{-\beta E_k}
\right)
\end{equation}
where
\begin{equation}
E_0 = -\frac{N}{2 \pi} \int_{0}^{\pi} dk \, \, E_k
\end{equation}
is the ground state energy.

{}From (2.1), (3.9) and  (4.7) we get
\begin{equation}
C= \frac{6 \beta \Gamma}{\pi^2} \int_{0}^{\pi} dk \, \ln\left( 1 +
e^{-\beta  E_k} \right)
\end{equation}
and although we do not have an analytic form for (4.9) we
can study its limits.

Observe that at the critical point, $\varrho =1$ ($J=2 \Gamma$),
we can rewrite (4.9) as
\begin{equation}
C=\frac{6 \beta \Gamma}{\pi^2} \sum_{n=1}^{\infty} \frac{(-1)^n}{n}
\int_{0}^{\pi/2} dk \, e^{-2 n \beta \Gamma \sin(k/2)}
\end{equation}
and therefore at the critical point $C$ depends on the temperature.
This occurs because we know that the continuum limit can only be
taken at low temperatures. Recall that $k$ is the lattice momentum. For
a system with lattice constant $a$, the momentum $q$ in laboratory units
is $q=k/a$. The continuum limit is the limit of $a \to 0$, keeping
the physical dimensionless temperature $t^{-1}= \Gamma a/T$ fixed. In
this limit, (4.10) becomes
\begin{equation}
\lim_{a \to 0} C=\lim_{a \to 0} \, \, \frac{6}{\pi^2 t} \sum_{n=1}^{\infty}
\frac{(-1)^n}{n} \int_{0}^{\frac{\pi}{2a}} dq e^{- \frac{2n}{t}
\sin\left(\frac{q \, a}{2}\right)} =\frac{1}{2}
\end{equation}
as expected from the results of conformal field theory.

We can also understand this result from the point of view of
section 3. Observe that at low temperatures and close to
the phase transition we can expand (4.3) around $k=0$ as
\begin{equation}
E_k = \Gamma \sqrt{(1-\varrho)^2 + \varrho \, k^2}
\end{equation}
which has the same form as (3.3) where the mass term (or the inverse of
the correlation length)
is proportional to $1-\varrho$ which vanishes at the
critical point. If now we let $k$ varies
between $0$ and infinity we recover the result of the former
section.

Observe that here the fact that the energy cuttoff ( {\it i.e.} the bandwidth
of the spectrum)
is finite is very important. We can not excite particles in the system
with an energy larger than the bandwidth and therefore we have
a temperature cut-off exactly as discussed in second section
for the case of constraint (2.24). This fact will result that
the $C$ function defined in (4.9) will decrease with temperature
for temperatures larger than the bandwidth. If we take the
limit that $\beta \Gamma << 1$ in (4.10) we easily see that
$C$ goes to zero in this limit.

We will pay attention to two limits, the limit where the we have
unbroken symmetry (the vaccum is invariant under the symmetry of
the Hamiltonian) and the limit of broken symmetry \cite{shenker}.
When $\varrho << 1$ ($J<<\Gamma$) the vaccum is an eigenstate
of $S^3$ where all spins are pointing up. From (4.9) is easy
to see that
\begin{equation}
C=\frac{6 \beta \Gamma}{\pi} \, \ln \left(1 + e^{-\beta \Gamma}
\right)
\end{equation}
and get a behaviour where $C$ increases monotonically in the
low temperature region and when reachs temperatures of order
of the bandwidth, namely $\Gamma$, $C$ goes monotonically to
zero (see Fig.2).

When $\varrho >> 1$ ($J>>\Gamma$) the vaccum is an eigenstate
of $S^1$ and it is degenerate (all the spins can be up or down),
that is, the symmetry is broken.
{}From (4.9) is trivial to show that
\begin{equation}
C=\frac{6 \beta \Gamma}{\pi} \, \ln\left(1 + e^{-\frac{\beta J}{2}}
\right)
\end{equation}
and therefore we see that $C$ increases monotonically from $T=0$ ($C=0$)
to temperatures of order $J$ and then decreases monotonically to
zero again. We see that the function $C$ in the low temperature
regime increases monotonically with the temperature as expected by
Zamolodchikov's theorem and the cut-off temperature,
as explained in (2.24), is $J$. This result can also be obtained from the
free energy of a {\it classical} one-dimensional Ising model. It is well
known that this trivial classical model has an infrared unstable fixed
point at $T=0$. We call this the classical zero temperature fixed point.
In this classical problem it is obvious that the free energy is not a
monotonic function of the temperature even though the correlation length
is indeed monotonic. Monotonicity is only present near the fixed point,
{\it i.e.} near $T=0$.

Anyway, even away from the critical point the monotonicity
properties of the $C$ function are consistent with Zamolodchikov's
theorem if we are in the region where the temperature is low
compared with the temperature cut-off of the problem, exactly
as considered in (2.24). Again the behaviour of the $C$ function
is non analitic in $\varrho$, this is a clear sign of problems
in one dimension.

\setcounter{equation}{0}

\section{The Non-Linear Sigma Model}

The non-linear sigma model is the described by the following
Lagrangean density
\begin{equation}
L=\frac{1}{2 g} \, \left( \partial_{\mu} \vec{n}(\vec{x},t)
\right)^2
\end{equation}
with $\mu=1,...,d+1$ and the constraint
\begin{equation}
\left(\vec{n}\right)^2=1
\end{equation}
where $\vec{n}$ in a vector with $N$ components and $g$ is the coupling
constant.

It can be shown that the Heisenberg antiferromagnet in the continuum limit
can be described by the non-linear sigma model ( see \cite{fradkin}
and references therein). This
model has been studied in the context of the High-Temperature
materials with the use the renormalization group method \cite{sudip}.
Here we will use the expansion $1/N$ in order to study its properties
and we will see that our results are equivalent
to those found in \cite{sudip} and \cite{taka}.

The thermodynamical partition function is written as
\begin{equation}
Z=\int D\vec{n}(\vec{x},\tau) \delta\left(\vec{n}^2-1\right)
e^{-\frac{S_0\left[ \vec{n} \right]}{g}}
\end{equation}
where $\tau$ is the imaginary time and
\begin{equation}
S_0=\int_{0}^{\beta}d\tau \, \int_{-\infty}^{+\infty}dx^d \,
L(\vec{n}(\vec{x},\tau))
\end{equation}
is the euclidean action with $\beta=1/T$ is the inverse of the
temperature.

Observe that in the low temperature limit we can extend the
integral (5.4) over the whole imaginary time axis and we
find a non-linear sigma model in $d+1$ dimensions with the
coupling constant $g$. In the high temperature limit the
upper limit in the integral in (5.4) is small and we can
rewrite the action as the action for the classical non-linear
sigma model in $d$ dimensions with a coupling constant given
by $g T$ \cite{sudip}.

We now decompose the field $\vec{n}$ in its components as
\begin{equation}
\vec{n}=(\sigma,\vec{\pi})
\end{equation}
where $\vec{\pi}$ is a vector with $N-1$ components.
Substituting (5.5) in (5.3) changing variables in the functional
integral as $\vec{\pi}=\sqrt{g} \vec{\phi}$ and tracing over $\vec{\phi}$
we get
\begin{equation}
Z=\int D\sigma(\vec{x},\tau) \int D\lambda(\vec{x},\tau) e^{-\frac{S\left[
\sigma,\lambda \right]}{g}}
\end{equation}
where we have introduced a representation for the delta function and
\begin{equation}
S=\int_{0}^{\beta}d\tau \int_{-\infty}^{+\infty} d^d x  \, \frac{1}{2}
\left[ \left(\partial_{\mu} \sigma\right)^2 + \lambda \left(\sigma^2
-1\right)\right] + \frac{N-1}{2} tr \ln \left(-\partial^{2}_{\mu} + \lambda
\right)
\end{equation}
where $tr$ is the trace.

Now rescale the parameters as
\begin{equation}
\tilde{g}=(N-1) g
\end{equation}
\begin{equation}
\sigma = \sqrt{(N-1)g} \, \, \tilde{\sigma}
\end{equation}
and rewrite (5.6) as
\begin{equation}
Z=\int D\tilde{\sigma} \int D\lambda \, \, e^{-(N-1) S_{ef}
 \left[\tilde{\sigma},
\lambda \right]}
\end{equation}
where the effective action is given by
\begin{equation}
S_{ef}= \int_{0}^{\beta} d\tau \int^{+\infty}_{-\infty} d^d x \,
\frac{1}{2} \left[ \left(\partial_{\mu} \tilde{\sigma} \right)^2
+ \lambda \tilde{\sigma}^2 - \frac{\lambda}{\tilde{g}}\right]
+ tr \ln \left(-\partial_{\mu}^{2} + \lambda \right)
\end{equation}
In the limit as $N \to \infty$ the solution of (5.10) can be obtained
by the saddle point equations
\begin{equation}
\left(-\partial_{\mu}^{2} + \lambda \right) \tilde{\sigma} = 0
\end{equation}
\begin{equation}
\tilde{\sigma}^2 = \frac{1}{\tilde{g}} - G_{\lambda}\left(\tilde{x},\tau;
\tilde{x},\tau\right)
\end{equation}
and
\begin{equation}
\left(-\partial_{\mu}^2 + \lambda \right) G_{\lambda}\left(\vec{x},\tau;
\vec{x'},\tau'\right)= \delta^d\left(\vec{x}-\vec{x'}\right) \delta\left(\tau-
\tau'\right)
\end{equation}
Assuming that $\tilde{\sigma}$ and $\lambda$ are constants we
can replace the equations above by
\begin{equation}
\lambda \sigma=0
\end{equation}
and
\begin{equation}
\sigma^2 = 1 - \frac{\tilde{g}}{\beta} \sum_{n} \int \frac{d^dk}{(2\pi)^d}
\frac{1}{k^2 + \omega_{n}^2 + \lambda}
\end{equation}
where we have used the periodic boundary conditions for the fields in the
imaginary time direction
\begin{equation}
\vec{\phi}\left(\vec{x},\tau\right)=\vec{\phi}\left(\vec{x},\tau + \beta\right)
\end{equation}
and therefore $\omega_n=\frac{2 \pi n}{\beta}$ with $n=0,\pm1,\pm2,...$.

We can proceed further and perform the sum in (5.16) \cite{grads}
\begin{equation}
\sigma^2 = 1 - \frac{\tilde{g} S_d}{2} \int_{0}^{\infty} dk \, k^{d-1}
\frac{\coth\left(\frac{\beta}{2} \sqrt{k^2 + \lambda}\right)}{\sqrt{k^2 +
\lambda}}
\end{equation}
where $S_d^{-1}=2^{d-1} \pi^{\frac{d}{2}} \Gamma(\frac{d}{2})$.

{}From (5.15) is easy to see that we have two phases in the problem, one
ordered
phase ($ \lambda = 0$ and $\sigma \not= 0$) and one disordered phase ($ \lambda
\not= 0$ and $\sigma = 0$). In the ordered phase we have $N-1$ Goldstone
bosons and one massive boson and in the disordered phase the full symmetry of
the system is recovered and the modes are massive.

Notice that expression (5.18) contains a ultra-violet divergence
($k \to \infty$)
at zero temperature in one and two dimensions which means that we
have to renormalize
the theory. In order to do so define the renormalization group transformations
\begin{equation}
\sigma = \sqrt{Z_0} \, \, M
\end{equation}
\begin{equation}
\lambda=\frac{Z_1}{Z_0} \, \Lambda^2 \, m^2
\end{equation}
\begin{equation}
\tilde{g}= Z_1 \Lambda^{1-d} \alpha
\end{equation}
\begin{equation}
T=\Lambda t
\end{equation}
\begin{equation}
U=\tilde{g} T= Z_1 \Lambda^{2-d} u
\end{equation}
where
\begin{equation}
u=\alpha t
\end{equation}
Here $M$, $m$, $\alpha$, $t$ and $u$ are dimensionless quantities
and $\Lambda$ is a scale with
dimensions of inverse of length ($\Lambda=1/a$, where $a$ is the lattice
spacing). Observe that the temperature $T$ is not renormalized since
it fixes the length of the manifold (the cilinder). In order to
achieve the renormalization in two dimensions we have to introduce
a new variable $U$, which is the high temperature coupling constant
as explained before.

It is easy to see that the theory is renormalized in any number of
dimensions if we choose
\begin{equation}
Z_0=Z_1
\end{equation}
and
\begin{equation}
\frac{1}{Z_0}=1+\frac{\alpha K_d}{2} \int_{0}^{\infty} dx \, x^{d-1}
\frac{\coth\left(\frac{\alpha}{2u} \sqrt{x^2+1}\right)}{\sqrt{x^2+1}}
\end{equation}

{}From (5.26) is possible to calculate the $\tilde{\beta}$-functions
of the model. Using (5.21), (5.22) and (5.23) we have
\begin{equation}
\tilde{\beta}_{\alpha}=\Lambda \left(\frac{\partial\alpha}
{\partial\Lambda}\right)_{g,T}
=(d-1)\alpha -\alpha \, \, \Omega(\alpha,u)
\end{equation}
\begin{equation}
\tilde{\beta}_{u}=\Lambda \left(\frac{\partial u}
{\partial \Lambda}\right)_{g,T}
=(d-2)u - u \, \, \Omega(\alpha,u)
\end{equation}
and
\begin{equation}
\tilde{\beta}_{t}=\Lambda \left(\frac{\partial t}
{\partial \Lambda}\right)_{g,T}=-t
\end{equation}
where
\begin{equation}
\Omega(\alpha,u)=\Lambda \left(\frac{\partial ln Z_0}
{\partial \Lambda}\right)_{g,T}
=\tilde{\beta}_{\alpha} \left(\frac{\partial ln Z_0}
{\partial \alpha}\right)+
\tilde{\beta}_{u} \left(\frac{\partial ln Z_0}{\partial u}\right)
\end{equation}
We can now solve the system formed by (5.27) and (5.28) using
expression (5.26). Although we can do it in general we will
choose $d=2$ since all integrals can be done explicitly.
For $d=2$ we find (the zero temperature integrals are done as in
\cite{ramond} in the context of $\epsilon$ expansion)
\begin{equation}
\tilde{\beta}_{\alpha}=\alpha-\frac{\alpha^2}{4 \pi} \,
\coth\left( \frac{\alpha}{2 u}\right)
\end{equation}
and
\begin{equation}
\tilde{\beta}_{u}=-\frac{\alpha u}{4 \pi} \, \coth\left(\frac{\alpha}
{2 u}\right)
\end{equation}
which agrees with the one loop result found in \cite{sudip}.
Furthermore, we can solve (5.18) explicitly and one finds
(use definition (5.24))
\begin{equation}
m = 2 t \, arcsinh\left(e^{\frac{2 \pi(M^2\alpha-1)}{\alpha t}}
\sinh\left(\frac{1}{2 t}\right)\right)
\end{equation}
which, together with
\begin{equation}
M \, m = 0
\end{equation}
completes the solution of the saddle point equations.

Notice that the fixed points of the model (where the the $\tilde{\beta}$-
functions vanish) are $u=0$, $\alpha=0$ and $u=0$, $\alpha=4\pi$ (or
simply, $(0,0)$ and $(0,4\pi)$). $(0,0)$ is a stable fixed point
and $(0,4\pi)$ is unstable.

Observe that at zero temperature in the ordered phase ($m=0$,
$M \not= 0$) we get
\begin{equation}
M^2=1-\frac{\alpha}{\alpha_c}
\end{equation}
where
\begin{equation}
\alpha_c=4 \pi
\end{equation}
is the critical dimensionless coupling constant which was expected from
the above analisys of the $\tilde{\beta}$-functions.
Observe that the expression (5.35) is only valid as $\alpha \leq \alpha_c$.
In the desordered phase ($m\not= 0$, $M=0$, $\alpha \geq \alpha_c$) we get
\begin{equation}
m=1-\frac{\alpha_c}{\alpha}.
\end{equation}

We can also solve exactly the equation (2.9)
for the running coupling constant
\begin{equation}
\alpha(t)= \alpha t_0 \left(t + \frac{\alpha t_0 \, t}{2\pi}
\, \, \ln\left(\frac{\sinh\left(\frac{1}{2t}\right)}{\sinh\left(
\frac{1}{2t_0}\right)}\right)\right)^{-1}.
\end{equation}

We can easily see that the dimensionless correlation length
as defined by (2.7) is $m^{-1}$ (use (5.33) in (2.9)).
Another possible way to achieve this result is to
expand the action (5.11) around the saddle point solutions
and recall the definition of the free energy
\begin{equation}
F=-\frac{1}{\beta} \ln Z
\end{equation}
we get that the finite term (which does not need to be regularized)
in the leading order term in $1/N$ is given by
\begin{equation}
\frac{F-E_0}{(N-1) V}= -\frac{S_d  \, 2^{\frac{d-1}{2}} \Gamma\left(
\frac{d}{2}\right)}{\pi^{1/2} \beta^{d+1}} \left(\beta \lambda^{1/2}
\right)^{\frac{d+1}{2}} \sum^{\infty}_{n=1} \frac{1}{n^{\frac{d+1}{2}}}
\, K_{\frac{d+1}{2}} \left(n \beta \lambda^{1/2}\right)
\end{equation}
The free energy only depends on the mass of the
$N-1$ modes, which is represented by $\lambda$.
Notice the resemblance with expression (3.6), the main difference
is that here the correlation length depends on the temperature.
Morover, comparing (5.40) with (2.1) we find the $C$ function,
\begin{equation}
C=\frac{S_d \, 2^{d-1} \Gamma\left(\frac{d}{2}\right)}{n(d) \, \pi^{1/2}}
\, \left(\frac{m(\alpha,t)}{t}\right)^{\frac{d+1}{2}} \, \sum_{n=1}^{\infty}
\frac{1}{n^{\frac{d+1}{2}}} K_{\frac{d+1}{2}}\left(n \frac{m(\alpha,t)}{t}
\right).
\end{equation}
Comparing (5.41) with (2.8) we conclude that $\gamma=m^{-1}$, as expected.

Now we want to know how $C$ changes with the
temperature. From (3.8) is easy to see that
\begin{equation}
\frac{dC}{d\left(\gamma t\right)} \geq 0
\end{equation}
and from (5.33) one finds
\begin{equation}
\frac{\partial \gamma}{\partial \alpha} \leq 0.
\end{equation}
Using our result (2.17) we arrive to the conclusion that the
sign of the derivative in (2.17) equals to minus the sign
of the beta function.

We can clearly distinguish two regions in the parameter space.
One of them has a positive beta function and is called \cite{sudip}
the classical renormalized region because the correlation
length diverges exponentially near $T=0$ as it does in the classical
version of the theory with the bare coupling constant replaced by
the renormalized one (see (5.33)); the second region has a negative beta
function
and is called quantum region. The border between these regions is given
by the zeros of $\tilde{\beta}_{\alpha}$, namely,
\begin{equation}
\alpha=\alpha_c \, \, \tanh\left(\frac{1}{2t}\right)
\end{equation}
The meaning of this border line is clear, we see that at low temperatures
the correlation length at this line equals the thermal wavelength (which
in our units is $1/T$). Therefore, in the classical region the correlation
length is always greater than the thermal wavelength which gives the
classical character of the region, in the quantum region the correlation
length is smaller than the thermal wavelength, as expected.

We finally conclude that if we start the flow from the classical
region our $C$ function will decrease with increase the temperature (since
$\tilde{\beta}_{\alpha}$ is positive) as far as it finds the border
line (5.44) and then, in the quantum region, it will start to
increase monotonically with the increase of the temperature (see Fig.3).

We see that the non monotonicity
of the $C$ is again related with the crossover between quantum and classical
behaviour. In one dimension the crossover is simply dictated by the
temperature in a trivial way, that is, any lattice leads to a bandwidth in
the spectrum of excitations which is a natural temperature cut-off.
This behaviour is implicit in the original assumptions of the
Zamolodchikov's theorem. In higher dimensions the existence of
phase transitions (even in the presence of a infinite bandwidth
as in the case of the Non-linear Sigma model) produces a region
in the parameter space where the behaviour is clearly classical,
that is, where the correlation length is much greater than
the thermal wavelength.

Now we can understand why the Zamolodchikov's theorem is
so powerful in one dimension in the continuum limit.
In $d=1$ we have only a fixed point at $(0,0)$ and therefore
there is no classical region in the parameter space. This is
also true for systems with relativistic dispersion relation
and although we can arbitrarily
choose a region in the parameter space where the correlation length
(independent of temperature) is smaller than the thermal wavelength
we do not have a phase transition, that is, the correlation
length does not blow up dinamically.

\section{Conclusions}

We have shown in this article that, in regions of the parameter
space (coupling constants and temperature) where quantum
fluctuations dominate, it is possible to define a $C$ function
(in any number of dimensions) which has the same properties
as the $C$ function defined in Zamolodchikov's $C$ theorem.
We showed that
in the regions where the system behaves classically
(where the correlation length is greater than the
thermal wavelength) monotonic properties are not to be expected.

We illustrate these issues in the context of several models.
We showed that free massive field theories have the mononicity properties
proposed in Zamolodchikov's theorem in any number of dimensions.
This result is due to the lack of a classical region in the parameter
space of the theory, exactly as in one-dimensional field theories.
Applying our ideas for the case of the Ising model in 1+1
dimensions we showed that the Zamolodchikov's theorem applies
in the region of low temperatures (as expected) and the fact
that we have a bandwidth (or temperature cut-off) implies
a crossover between classical and quantum behavior.

In the case of the Non-linear Sigma model in
2+1 dimensions we found
that in regions where the correlation length is greater than
the thermal wavelength (that is, close to a phase transition)
the thermodynamic functions exhibit classical behavior and consequently
the function $C$ is not
a monotonic function of the temperature. Nevertheless, in regions
where the system is essentially quantum mechanical our
function $C$ does increase
monotonically with the temperature, a result similar to the Zamolodchikov's
$C$-theorem. In this way we see that $C$ can be used to describe the
crossover between classical and quantum behaviour.

Since our approach is thermodynamical, it is possible to
extract these behavior from experiments which measure the specific
heat of the system under consideration.

\section{Acknowledgments}

We would like to thank  J.~M.~P.~ Carmelo  and Andrea Cappelli for
useful discussions. A.~H.~C.~N thanks  CNPq (Brazil) for a fellowship.
This work was supported in part by NSF
Grant No.DMR91-22385 at the University of Illinois at Urbana-Champaign.


\newpage

\begin{thebibliography}{50}
\bibitem{bela} A.A.Belavin, A.M.Polyakov and A.B.Zamolodchikov,
Nucl.Phys.{\bf B241}, 333,(1984).
\bibitem{cardy} For review articles see, for example, J.L.Cardy,
in Field, Strings and Critical Phenomena, (Les Houches 1988), ed.
E.Brezin and J.Zinn-Justin (North Holland, 1990).
\bibitem{Zamol} A.B.Zamolodchikov, Pis`ma Zh.Eksp.Teor.Fiz. {\bf 43},
565,(1986);[JETP Lett. {\bf 43},730,(1986).]
\bibitem{cappelli1} A.Cappelli, D.Friedan and J.I.Latorre, Nucl.Phys.
{\bf B352},616,(1991).
\bibitem{cappelli2} A.Cappelli, J.I.Latorre and X. Vilasis-Cardona,
Nucl.Phys. {\bf B376},510,(1992).
\bibitem{boya}D.Boyanovsky and C.M.Naon, Rev.Nuo.Cim {\bf 13}, N.2,(1990).
\bibitem{zinn}J.Zinn-Justin, Quantum Field Theory and Critical Phenomena,
(Claredon Press, Oxford, 1989).
\bibitem{reif} F.Reif, Fundamentals of Statistical and Thermal Physics,
(McGraw-Hill, New York, 1965).
\bibitem{ludwig} A.W.W.Ludwig and J.L.Cardy, Nucl.Phys.{\bf B285},
687,(1987).
\bibitem{grads} I.S.Gradshtein and I.M.Ryzhik, Tables of Integrals,
Series and Products, (Academic Press, London, 1980).
\bibitem{affleck} I.Affleck, Phys.Rev.Lett.{\bf 56},746,(1986).
\bibitem{pines} D.Pines and P.Nozieres, Theory of Quantum Liquids,
 Vol.1, (Addsion-Wesley, Redwood City, 1989).
\bibitem{lieb} E.Lieb, T.Shultz and D.Mattis, Ann.Phys.(N.Y.)
{\bf 16}, 407, (1961).
\bibitem{pfeuty} P.Pfeuty, Ann.Phys.(N.Y.) {\bf 57}, 79, (1970).
\bibitem{shenker} H.Shenker in Recent Advances in Field Theory
and Statistical Mechanics, Les Houches (1982).
\bibitem{fradkin} E.Fradkin, Field Theories of Condensed Matter
Physics,(Addison-Wesley, Redwood City, 1991).
\bibitem{sudip} S.Chakravarty, B.I. Halperin and D.R.Nelson,
Phys.Rev.{\bf B 39}, 2344, (1989).
\bibitem{taka}T. Yanagisawa, Phys.Rev.Lett. {\bf 68},
1026,(1992).
\bibitem{ramond}P. Ramond, Field Theory: A Modern Primer, (Addsion-
Wesley, Redwood City, 1989).
\end{thebibliography}
\end{document}